\documentclass[preprint,12pt]{elsarticle}
\pdfoutput=1
\usepackage{epsfig}
\journal{Physica A}

\begin{document}

\title{Dynamics of Snoring Sounds and Its Connection with Obstructive Sleep Apnea}

\author{Adriano M. Alencar}
\address{Instituto de Fisica, Universidade de S\~ao Paulo, SP, Brazil}
\ead{adralencar@if.usp.br}

\author{Diego Greatti Vaz da Silva, Carolina Beatriz Oliveira}
\address{Sleep Laboratory, Pulmonary Division, Heart Institute (InCor), Faculty of Medicine, University of Sao Paulo, SP, Brazil}

\author{Andr\'e P Vieira}
\address{Instituto de Fisica, Universidade de S\~ao Paulo, SP, Brazil}

\author{Henrique T Moriya}
\address{Biomedical Engineering Laboratory of University of S\~ao Paulo, SP, Brazil}

\author{Geraldo Lorenzi-Filho}
\address{Sleep Laboratory, Pulmonary Division, Heart Institute (InCor), Faculty of Medicine, University of Sao Paulo, SP, Brazil}

\date{\today}

\begin{abstract}
Snoring is extremely common in the general population and when irregular may
indicate the presence of obstructive sleep apnea. We analyze the overnight
sequence of wave packets --- the snore sound --- recorded 
during full polysomnography in patients
referred to the sleep laboratory due to suspected obstructive sleep apnea. We
hypothesize that irregular snore, with duration in the range between 10 and 100 seconds,
correlates with respiratory obstructive events. We find that the number of
irregular snores --- easily accessible, and quantified by what we call the snore 
time interval index (STII) ---
is in good agreement with the well-known
apnea-hypopnea index, which expresses the severity of obstructive sleep apnea
and is extracted only from polysomnography. 
In addition, the Hurst analysis of the snore
sound itself, which calculates the fluctuations in the signal as a function of
time interval, is used to build a classifier that is able to distinguish 
between patients with no or mild apnea and patients with moderate or severe apnea. 
\end{abstract}


\maketitle

\section{Introduction}

Temporal patterns of natural events can be used to detect hidden dynamics
\cite{vandewalle-2001}. In fact, this connection dates back to Hurst's work\,\cite{hurst51}
on the long-term memory (i.e. the autocorrelation) of
a time series. This approach has been used to analyze respiratory sound events,
and for instance the time interval between consecutive crackling sounds,
generated when a collapsed airway opens, has been related to both lung structure
and surface properties of the lung coating fluids\,\cite{alencar-2001,alencar-2003}.
The dynamics of various other physiological signals has been investigated 
with the help of tools derived from information theory and statistical
mechanics (see e.g.\,\cite{ishizaki-2008,acosta-2011} and references therein).

Snoring is perhaps the best known respiratory sound, and is  characterized by a
loud sound with frequency content between 20 and 300\,Hz. It is caused by the
vibration of soft tissues from the upper airway, including soft palate, uvula,
tonsils, tonsillar pillars, base of tongue, lateral pharyngeal walls and mucous
membranes \cite{fairbanks-1994}. Snoring occurs during sleep, when the
pharyngeal muscles relax and may block the airway. 
Snoring is common in the general population, and does not necessarily
indicate a disease. In contrast, irregular snoring is one of the hallmark signs
suggestive of Obstructive Sleep Apnea (OSA). OSA is the most common
sleep-disordered breathing, and is characterized by repetitive events of upper
airway obstruction during sleep, causing total or partial cessation of airflow
(apneas and hypopneas, respectively)\,\cite{flemons-2002}. 

A respiratory event is defined as apnea or hypopnea if its duration exceeds 10
seconds. Apnea is characterized by cessation of airflow, while hypopnea is
defined as a substantial reduction -- of at least 50\% -- in airflow. OSA is a
major public-health problem due to the high prevalence in the adult population
(ranging from 4\% to 10\%), and due to the poor outcome when not recognized and
treated. OSA is associated with recurrent asphyxia, fragmented sleep and
generation of negative intrathoracic pressure during futile efforts to 
breath\,\cite{drager-2011}.
Although the patient is by and large unaware of these recurrent respiratory
episodes during sleep, the consequences are multiple and may include excessive
daytime sleepiness, fatigue, poor cognitive function and quality of life, as well
as increased risk of motor vehicle accidents, metabolic and cardiovascular
diseases. Full polysomnography is the gold standard method for OSA diagnosis,
and the most important index expressing the severity of OSA is the
apnea-hypopnea index (AHI), which is the average number of apnea or hypopnea
events per hour. Sleep apnea is usually classified into mild, moderate or
severe, depending on whether the AHI lies between 5 and 15 (mild), 15 and 30
(moderate), or above 30 events per hours (severe)\,\cite{ruehland-2009}. 
The major limitation of
polysomnography is that it is an expensive and laborious test, which requires an
overnight sleep in a specialized laboratory under the supervision of a
sleep technician. Therefore, the development of alternative and simpler methods
for the diagnosis of OSA is a priority\,\cite{chervin-2000}.

Irregular snoring and witnessed apneas are the major signals indicating
OSA and there is an increasing amount of research
trying to relate OSA with snoring patterns. 
Most
methods in the literature try to establish a correlation between OSA and
either the number of snore
events or the wave features of the snore such as intensity, sound
frequency, number of snore events, spectral density, or a combination of these
features~\cite{mainmon-2010,ghaemmaghami-2010,ng-2009}. 
Recently, Cavusoglu \emph{et al.} \cite{cavusoglu-2007,cavusoglu-2008} proposed
the use of sequential properties of snoring episodes for OSA identification. 
They built sequences of snoring episode duration, snoring episode time
separations and average snoring episode power, and were able to show that,
using statistical properties of these sequences to measure snore regularity,
OSA patients exhibit a greater degree of irregularity as compared with simple
snorers. 
Following a related path, in this work we focus on the dynamics of irregular
snoring episodes, and define a new Index, called the Snore Time Interval 
Index (STII), showing a strong positive correlation with the well-known
apnea-hypopnea index. Furthermore, by a fluctuation analysis of the 
snore sound, through Hurst's R/S analysis, we build an automated classifier
which is able to distinguish between patients with no or mild OSA and 
patients with moderate or severe OSA.
In a similar spirit, a number of studies focused on the study
of dynamic properties of sleep electroencephalogram (EEG) signals and their
relation to OSA\,\cite{lee-2004,rajendra-2005,zhang-2009}. Here we work directly with acoustic signals, whose
acquisition, in contrast with EEG signals, requires very simple equipment and
no direct contact with the patients.

The remaining of this paper is organized as follows. In section \ref{sec:methods}
we describe details regarding data acquisition and processing, and introduce
the mathematical definitions used in this work. In section \ref{sec:results} we
present the results, and the paper ends in section \ref{sec:conclusions}, in
which we provide a discussion and our conclusions.

\begin{figure}
\includegraphics[width=0.8\columnwidth]{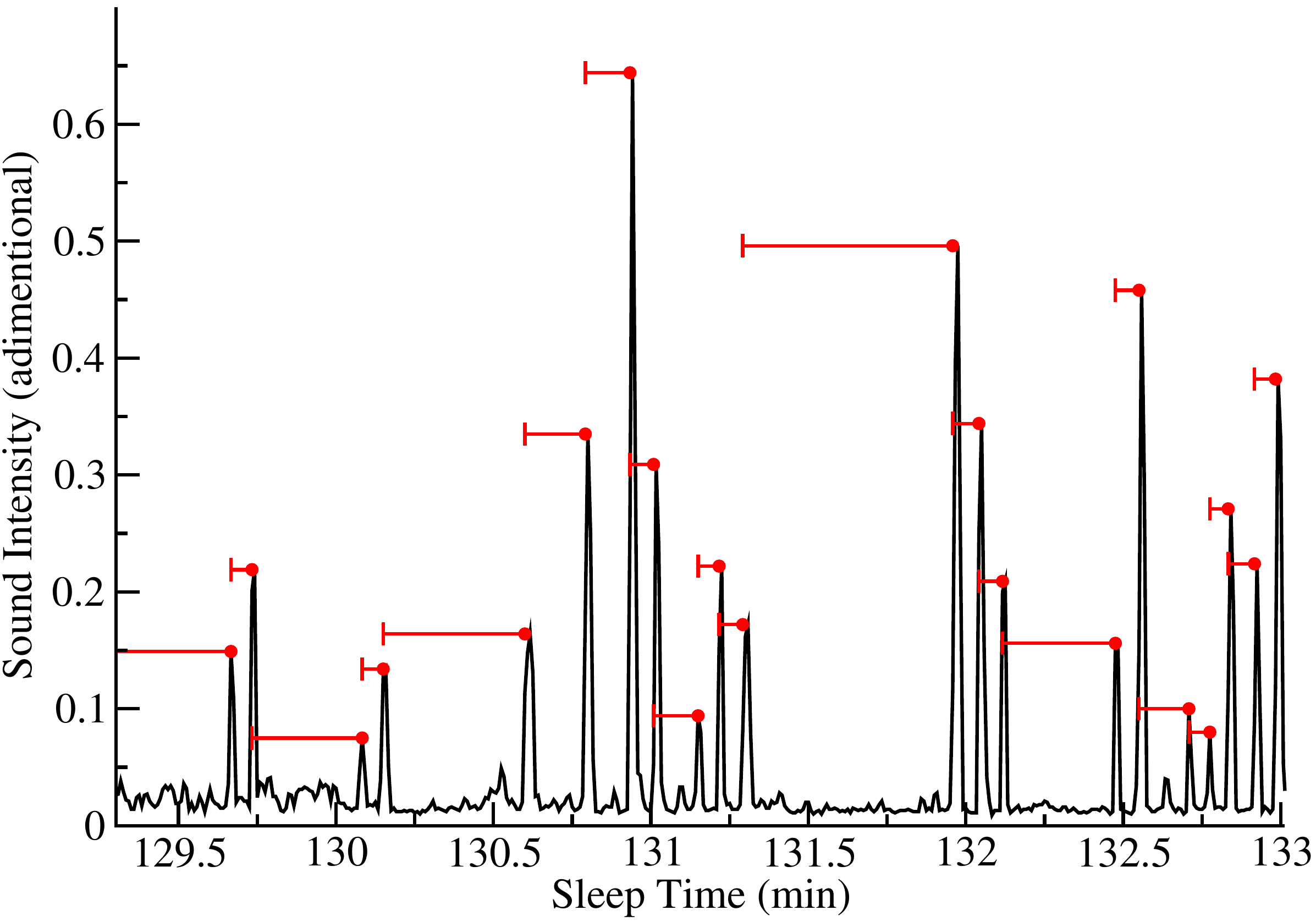}  
\caption{ Sound intensity versus time, showing the time intervals between
consecutive detected snore events. 
\label{fig:snd}}
\end{figure}

\section{Methods}
\label{sec:methods}

\subsection{Experimental and pre-processing details}

We analyzed snore records from patients referred to the sleep laboratory due to
suspected OSA. The patients slept in a bed at the Sleep Laboratory of the Heart
Institute at Clinics Hospital in S\~ao Paulo, Brazil, and we recorded the sounds
in the room while the patients were simultaneously
evaluated by a full polysomnography exam.

The population of 17 patients for which sound and polysomnography data are 
available was characterized by a male/female ratio of 11/6, average age 50$\pm$10 years, 
average body mass index 30.7$\pm$6.7\,kg/m$^2$, average 
apnea-hypopnea index 25.5$\pm$22.5\,events/h. 
Out of the 17 patients in the present study, 4 suffered from mild OSA, 
6 from moderate OSA, and 7 from severe OSA.

The sound-recording system is based on a 16-bit audio codec (PCM 2901, Texas
Instruments, USA) with a 4th order low-pass Butterworth analog filter (cutoff
frequency of 5\,kHz) and a 4th order high-pass Butterworth analog filter (cutoff
frequency of 20\,Hz). The audio codec is responsible for the acquisition
(sampling frequency of 44.1\,kHz) of the sound captured by the microphones, and
the transmission of the recorded data to a laptop computer by a USB serial
input. The sound was captured by two microphones located at the headboard of the
bed, at a distance of 90\,cm from each other. 

The microphones acquired the sound $S(t)$ emitted by the patient plus the
ambient noise, with a total time $T$, and recorded the data into a microcomputer
for later analysis. Since the frequency content of the snore signal is
predominantly between 20 and 300\,Hz, the first step of the analyses was to
digitally filter the signal, reducing the noise ratio. We found that the most
suitable filter for dismissing the unwanted ambient noise while still keeping a
strong snore signal was a 1024th-order band pass filter between 80 and 300\,Hz. 

Following the filtering, we square the filtered signal $S_f(t)$ and sum over all
$[S_f(t)]^2$ in a time window $\Delta t$, with a window overlap of 50\% between
consecutive windows, building a new series of sound intensities $I_m$,
\begin{equation}
I_m=\sum_{t=\frac{m}{2}\Delta t}^{(\frac{m}{2}+1)\Delta t}[S_f(t)]^2,
\end{equation}                            
with $m$ going from zero to $2(T/\Delta t-1)$, i.e., the series $\{I_m\}$
has a total of $2(T/\Delta t)-1$ elements. We choose $\Delta t = 1\,\mathrm{s}$,
since the normal breathing period is approximately 5 seconds, and we found that
this discretization gives us a good overall precision for our study while
keeping the signal less rough. We notice that the new series still exhibits
considerable background noise, most of it coming from air-conditioning and
electronic devices used by the Sleep Laboratory. From a region in the
series without snore, we could determine a threshold $I_0$ below which all
sounds are regarded as ambient noise, and then postulate that a group of points
in our new series defines a single snore event when all $I_m$ in the group lie
above the threshold $I_0$. According to this definition, the $j$th snore event
starts at point $m(j)$ and ends at the first point after $m(j)$ for which the
corresponding intensity is less than $I_0$. Thus, in time units the $j$th snore
event will start at  $(1/2)\Delta t\,m(j)$. We then define the time interval
between consecutive snores as the time from the beginning of one snore event to
the beginning of the next one, $\delta t_j = (1/2) \Delta t\,(m(j)-m(j-1))$, as
illustrated in figure\,\ref{fig:snd}.

At the moment, it is necessary to adjust the value of $I_0$ for each patient, 
considering
the intensities of the snore and the ambient noise. We built a software which is
able to read the sound data and process the signal, giving us a log file
containing the recording time in hours, minutes and seconds, the average, median
and standard deviation of the $\delta t_j$, as well as the maximum and minimum values 
of $\delta t_j$.

As breathing, snoring is an irregular phenomenon and it is an indicative of
instability in the upper airways. Snore usually persists at each breathing cycle
until one of three situations occurs:  (i) the obstruction momentarily
disappears; (ii) there is a partial collapse of the upper airways, restricting
the airflow and leading to hypopnea; (iii) the upper airway collapses,
impairing the airflow and leading to an apnea event. We assume that in case (i)
this momentary situation will persist for more than 100 seconds, and in both
cases (ii) and (iii), after the patient restores normal breathing, (s)he will
either get back to snoring or make a noise which will cross the ambient noise
threshold $I_0$. 

We then hypothesize that in a person that snores but does not have sleep apnea
disorder, most snore time intervals $\delta t$ will be less than 10\,s or
much larger than 100\,s. On the other hand, for a patient with OSA we should find
a lot of snore time intervals between 10\,s and 100\,s, which in fact we found.
Thus, in our analyses we focus on the range of time intervals $\delta t$ between
10\,s and 100\,s. 
 
\subsection{The Snore Time Inverval Index}

The Apnea-Hypopnea Index (AHI) is the classical index to determine if a person
has sleep apnea, and is only fully characterized when a person spends a night
of sleep in a sleep laboratory wired with electrodes and using an air mask to
measure air flow. Along the night a specialist medical doctor goes through the
data and manually counts the number of apnea and hypopnea events $N_{AH}$
occurring during the whole night. The AHI index is then evaluated as
\begin{equation} 
AHI = N_{AH} / T,
\end{equation} 
where $T$ is the number of sleep hours. 

From the time intervals $\delta t$, defined in the previous subsection, 
we introduce what we call the Snore
Time Interval Index (STII), which is related to the number $N_{\delta t}$ of
snore time intervals for which $10\,s < \delta t < 100\,s$, being explicitly
calculated as
\begin{equation} 
STII = N_{\delta t} / T.
\end{equation} 
As we will show in section \ref{sec:results}, there is a strong positive
correlation between AHI and STII.

\subsection{A classifier based on Hurst's R/S analysis}

Along the lines of \cite{vieira10}, we also built an automated classifier
from a fluctuation analysis of the intensities $I_m$, based on Hurst's $R/S$
method (see e.g. \cite{feder88}), which was introduced as a means of measuring
memory effects in a time series. The $R/S$ method involves dividing a time series
into intervals of size $\tau$, within which a rescaled range is calculated as
follows. For the $k$th interval, which we denote by $L_k$, we first determine
the average of the time series,
\begin{equation}
z_{\tau,k} = \frac{1}{\tau} \sum_{m\in L_k} I_m,
\end{equation}
which is used to define a cumulative deviation as
\begin{equation}
Z_{m,k} = \sum_{j=\ell_k}^{m}(I_j-z_{\tau,k}),
\end{equation}
$\ell_k$ labeling the left end of $L_k$. From this cumulative deviation, we then
extract a range,
\begin{equation}
R_{\tau,k} = \max_{m \in L_k}Z_{m,k}-\min_{m \in L_k}Z_{m,k},
\end{equation}
which is divided by the standard deviation of the original series,
\begin{equation}
S_{\tau,k} = \sqrt{\frac{1}{\tau} \sum_{m \in L_k} (I_m-z_{\tau,k})^2},
\end{equation}
to yield the rescaled range $R_{\tau,k}/S_{\tau,k}$ of a single interval.
Repeating this procedure for all intervals $L_k$, we obtain the average rescaled
range
\begin{equation}
\rho(\tau) = \frac{1}{n_{\tau}} \sum_{k} \frac{R_{\tau,k}}{S_{\tau,k}},
\end{equation}
where $n_{\tau}$ denotes the number of nonoverlapping intervals of size 
$\tau$ that can be fit into the original series.

When applied to a time series generated by a single dynamics, such as in fractional 
Brownian motion, the rescaled-range analysis yields a function $\rho(\tau)$
following a power-law with a so-called Hurst exponent $H$, which gauges
memory effects on the dynamics. Different values of $H$ indicate different
underlying dynamics, and this has been used in a number of situations
to distinguish between the systems producing the time series. 
Especifically, this approach can differentiate between healthy and
diseased subjects as regards cardiac \cite{goldberger2002}, 
neurological \cite{hausdorff2000,hwa2002},
and respiratory function \cite{peng2002}. However, as shown in the next Section,
this characterization in terms of a single parameter (the Hurst exponent)
fails in the case of OSA, but the $\rho(\tau)$ curves still hide
useful information about the patients conditions. In order to extract
that information, we resort to pattern-classification tools.

Given a set $\{\tau_j\}=\{\tau_1,\tau_2,\ldots,\tau_d\}$ of choices for 
the interval size $\tau$, the corresponding set
${\rho(\tau)}$ measures the average relative fluctuation of the time series as a
function of $\tau$, and this can be seen as a vector in a multidimensional
space,
\[
 \mathbf{x}=\left(\rho\left(\tau_1\right),\rho\left(\tau_2\right),
 \ldots,\rho\left(\tau_d\right)\right)^t,
\]
in which $t$ indicates the vector transpose.
The set of all vectors obtained by the $R/S$ analysis can then be fed into
standard algorithms from the pattern classification literature. Here, we employ
a variant of the Karhunen-Lo\`eve (KL) transformation (see 
\cite{webb02}), which can be seen as an improvement over principal-component
analysis, implementing supervision and relying on the compression of discriminatory
information contained in the class means.

In order to build a classifier whose performance is statistically
significant, we split the intensity signals from each patient into subsignals
containing 1024 points (corresponding to 512 seconds, an amount of time during
which many snore events can happen), obtaining a total of 936 subsignals for all
patients. The $R/S$ analysis was performed for values of the time window $\tau$
corresponding to the nearest integers obtained from the powers of 2$^{1/4}$ from
4 to 1024 points (with a total of 33 values). 

The first step of the KL transformation involves projecting the feature vectors $\mathbf{x}_i$
($i$ labelling the vectors obtained from different sound samples)
onto the eigenvectors of the within-class covariance matrix $\mathbf{S}_{W}$,
defined by 
\begin{equation}
\mathbf{S}_{W}=\frac{1}{N}\sum_{k=1}^{N_{C}}\sum_{i=1}^{N_{k}}y_{ik}(\mathbf{x}_{i}-
\mathbf{m}_{k})(\mathbf{x}_{i}-\mathbf{m}_{k})^{T},
\end{equation}
where $N_{C}$ is the number of different classes, $N_{k}$ is the
number of vectors in class $k$, and $\mathbf{m}_{k}$ is the average
vector of class $k$. The element $y_{ik}$ is equal to one if $\mathbf{x}_{i}$
belongs to class $k$, and zero otherwise.
The resulting
vectors are then rescaled by a diagonal matrix built from the eigenvalues $\lambda_{j}$
of $\mathbf{S}_{W}$. In matrix notation, this operation can be written
as \begin{equation}
\mathbf{X}^{\prime}=\mathbf{\Lambda}^{-\frac{1}{2}}\mathbf{U}^{T}\mathbf{X},
\end{equation}
in which $\mathbf{X}$ is the matrix whose columns are the training
vectors $\mathbf{x}_{i}$, $\mathbf{\Lambda}=\mathrm{diag}(\lambda_{1},\lambda_{2},...)$,
and $\mathbf{U}$ is the matrix whose columns are the eigenvectors
of $\mathbf{S}_{W}$. Finally, 
the resulting vectors are projected onto the eigenvectors of the between-class covariance
matrix $\mathbf{S}_{B}$, 
\begin{equation}
\mathbf{S}_{B}=\sum_{k=1}^{N_{C}}\frac{N_{k}}{N}(\mathbf{m}_{k}-\mathbf{m})(\mathbf{m}_{k}-\mathbf{m})^{T},
\label{eq:sb}
\end{equation}
where $\mathbf{m}$ is the overall average vector. The full transformation
can be written as 
\begin{equation}
\mathbf{X}^{\prime\prime}=\mathbf{V}^{T}\Lambda^{-\frac{1}{2}}\mathbf{U}^{T}\mathbf{X},
\end{equation}
$\mathbf{V}$ being the matrix whose columns are the eigenvectors
of $\mathbf{S}_{B}$ (calculated from $\mathbf{X}^{\prime}$).

\section{Results\label{sec:results}}

\begin{figure}
\includegraphics[width=0.8\columnwidth]{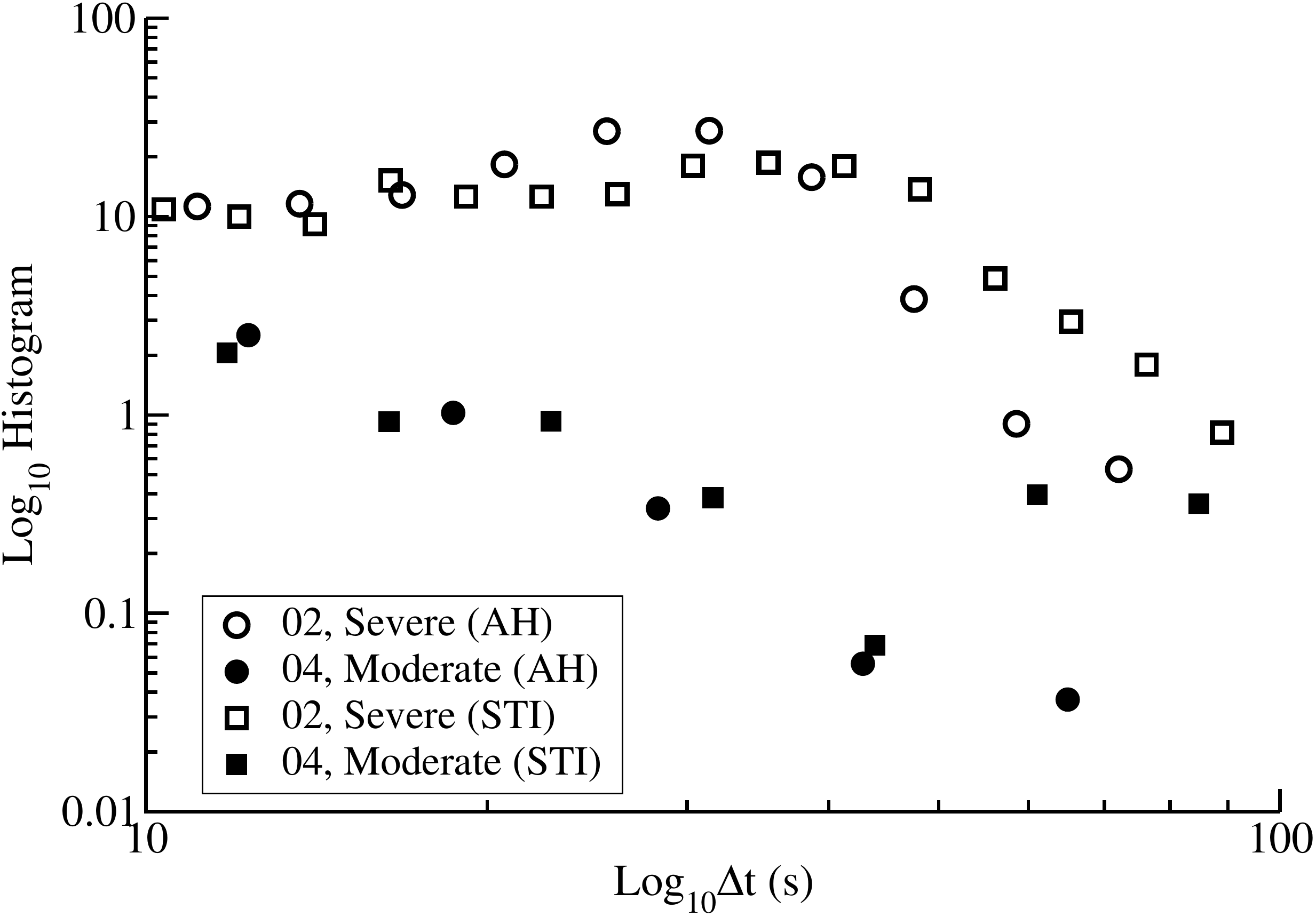}  
\caption{The histogram of time intervals $\Delta t$ between consecutive snores
(square) and the histogram of apnea-hypopneia duration (circles) in seconds. We
compare two extreme cases, filled and open symbols respectively for moderate and
severe OSA. 
\label{fig:aih-stii-histogram}}
\end{figure}

\begin{figure}
\includegraphics[width=0.8\columnwidth]{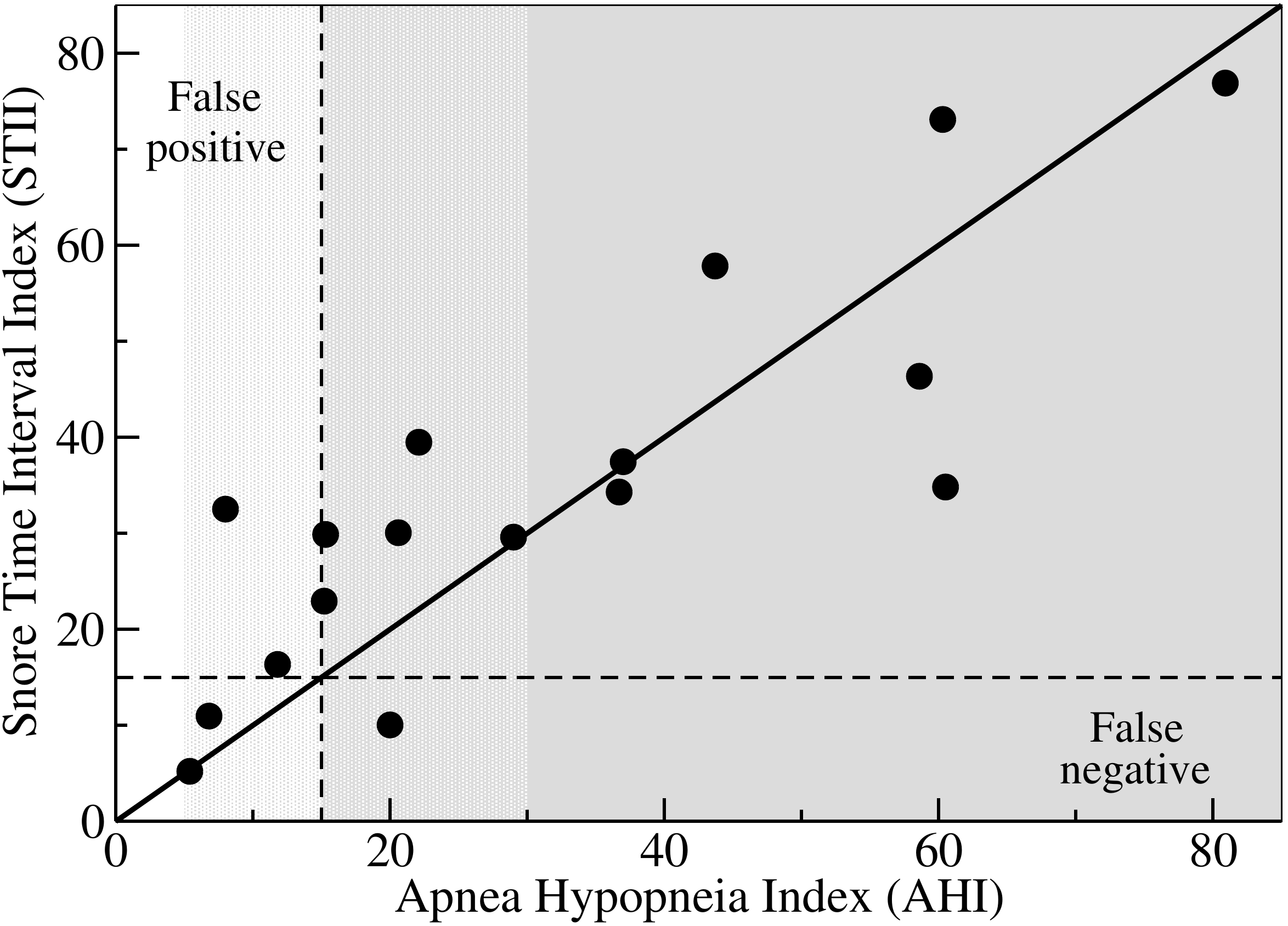}  
\caption{Snore Time Interval Index versus Apnea Hypopneia Index (AHI), the latter defined
by the polysomnography for 17 patients. The straight line is a line with slope
1 to highlight the correlation between the two indexes (correlation coefficient of 0.841, with
significance $t=6.02$, indicating a probability of less than $0.1\%$ for the null hypothesis
of uncorrelated data). 
As defined by the AHI value, the dark grey
region corresponds to severe OSA, the light grey region to moderate
OSA, the lightest grey region to mild OSA, and the white region to
normality. The dashed lines delimit the boundaries of the regions
where the STII leads to false positive (upper left rectangle) or 
false negative (lower right rectangle) results,
according to the classification in no/mild versus moderate/severe
OSA.
\label{fig:aih-stii}}
\end{figure}

As shown in figure\,\ref{fig:aih-stii-histogram}, there is a remarkable agreement 
between the histogram of the duration of apnea or hypopnea events and the 
histogram of snore time intervals. Moreover, our results reveal a strong
positive correlation between the apnea-hypopnea index and the snore-time-interval
index, as depicted in figure\,\ref{fig:aih-stii}. 
Notice that, assuming a perfect correlation between AHI and STII when it comes
to classifying OSA into no/mild versus moderate/severe, the STII yields only 
one false negative and
two false positive results (false positives being usually more tolerable mistakes).

\begin{figure}
\includegraphics[width=0.8\columnwidth]{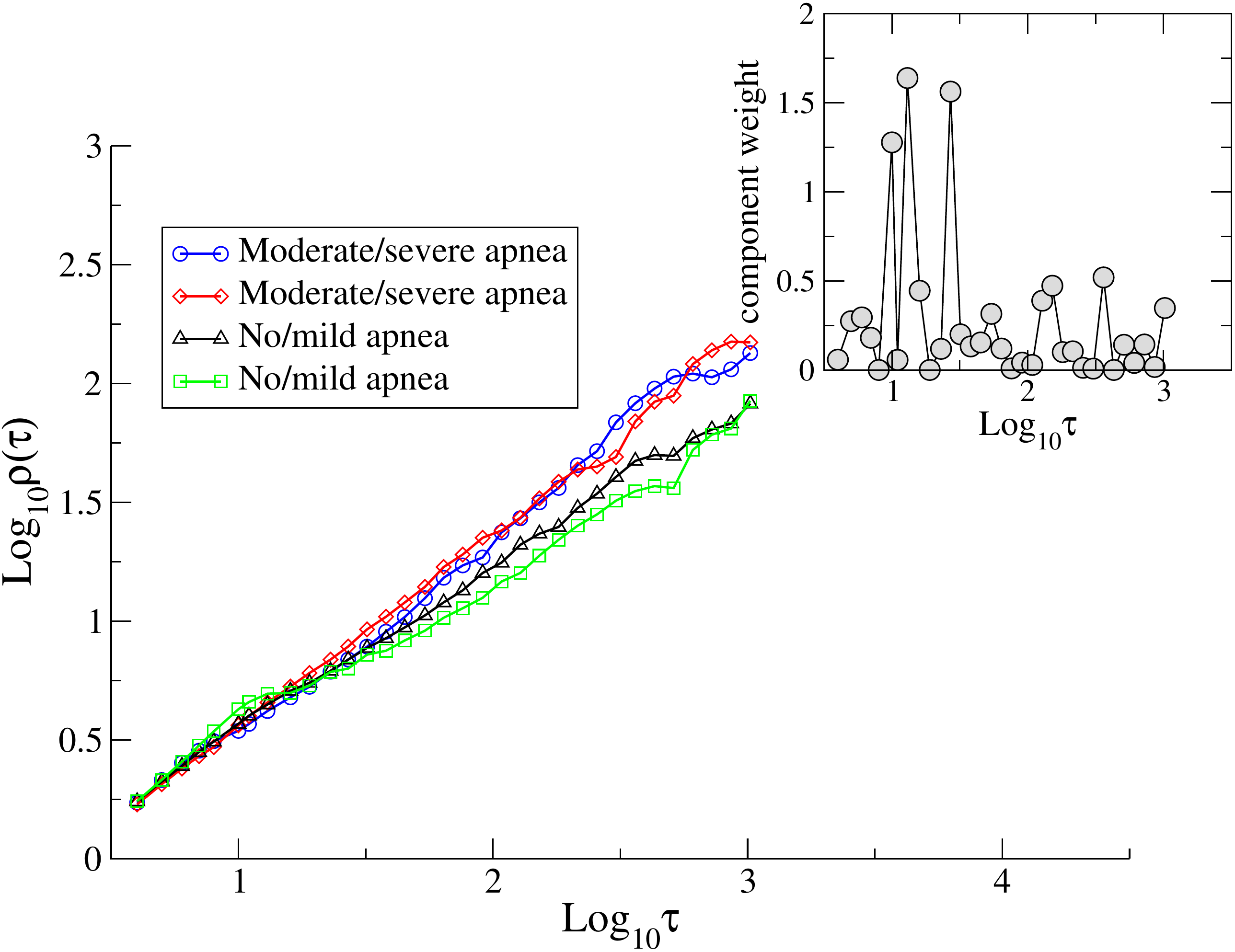}  
\caption{Main plot: examples of curves obtained by Hurst's $R/S$ analysis of the snore
sounds, corresponding to the average relative fluctuation of the signal, $\rho$,
versus the size of the time window, $\tau$, in double-logarithmic scale. 
Sound samples from 4 different patients were used, two with no or mild apnea,
and two with moderate or severe apnea. Inset: Weights of the components
of the Karhunen-Lo\`eve transformation vector corresponding to the 
different sizes of the time window. Projection of the fluctuation vector
of a given sound sample onto the Karhunen-Lo\`eve vector determines
the class to which the sample is attributed. 
\label{fig:rs}}
\end{figure}

Figure \ref{fig:rs} shows examples of curves obtained by Hurst's $R/S$ analysis
of sound samples taken from four different patients. Estimates of the Hurst
exponents of the different curves for all patients, obtained by linear fits of the log-log curves,
yielded results ranging between $0.7$ and $0.8$, with no significance correlation
to the patients condition. As described in the previous
section, the Hurst curves, interpreted as feature vectors, were then 
fed into the automated
classifier based on the KL transformation. The classifier was 
trained to recognize feature vectors of sound samples
coming from two categories: (a) patients with no or mild OSA; (b) patients
suffering from moderate or severe OSA. The training was performed by randomly
selecting 748 out of the available 936 subsignals (about 80 percent of the total), 
and using the remaining signals for testing. During the training step, the mean
transformed vectors of each class are extracted, and the (transformed) testing
vectors are assigned to the class corresponding to the nearest mean transformed
vector. With only two possible classes, the transformation matrix effectively reduces
to a single line, i.e. a transformation vector.\footnote{This is a consequence of the fact
that the subspace defined by the relative positions of the average vectors of 
the two classes, entering the definition of $\mathbf{S}_B$ in (\ref{eq:sb}), 
is one-dimensional.} The inset in figure \ref{fig:rs} shows, for
a typical transformation vector, the (unnormalized) weights of
the components corresponding to different sizes of the time window. The peaks in the curve
indicate those components which are most relevant for the decision
process. Notice the existence of large peaks around $\tau=20$ ($log\tau\simeq 1.3$),
corresponding to a time interval of $10,$s, a time interval characteristic of many
snore events in patients suffering from OSA.

This procedure was repeated for 100 distinct choices of training and testing
vectors. Based on each subsignal individually, the average classification error
during the testing step was 37\% for category (a) and 28\% for category (b).
However, if one extracts a classification from a majority rule of the
corresponding results for all subsignals belonging to a given patient, 16 of the
17 patients are correctly assigned to their respective categories.
Interestingly, the misclassified patient has an AHI of 15.2, and thus
lies on the borderline between categories.

\section{Conclusions\label{sec:conclusions}}

To summarize, in the present study we propose a simple and accurate method to 
identify OSA based on the time intervals between snore events, and the average 
number of
snore events per unit time, as revealed by the intensity of
the sound. 
Our method of analyzing snore correlated remarkably well with the AHI
derived from polysomnography, which is the gold standard method for OSA diagnosis. 
Moreover, an automated classifier based on the $R/S$ analysis of the snore
intensities
was able to correctly identify patients with moderate to severe OSA. 

Of course, our study has some limitations. Firstly, we tested this algorithm in a
relatively small sample of patients referred to the sleep laboratory under
controlled conditions. We have no means to predict at this stage how this method
will perform in other settings and populations. However, it must be stressed
that the physical conditions of the laboratory are probably similar to most
bedrooms. We acknowledge that due to the simplicity of our snore detection
scheme, some snore events may not be detected or some loud sound may be detected
as a snore. However, these false or undetected snore events are expected to
represent a small fraction of all events, and thus will be washed out
in the statistics. We predict that more sophisticated
methods to detect snores may improve the snore-time-interval index. 
Nevertheless, our work represents a first step towards establishing guidelines 
for snore classification and revealing a correlation between snoring events 
suspected of sleep-disordered breathing and the AHI, goals manifested in a recent 
medical review on the acoustics of snoring\,\cite{2010-pevernagie}.

\section{Acknowledgments}
This work was supported by Funda\c c\~ao de Amparo \`a Pesquisa do Estado de S\~ao
Paulo (FAPESP) and Conselho Nacional de Desenvolvimento Cient\'\i fico e Tecnol\'ogico (CNPq).

\section*{References}

\bibliographystyle{elsarticle-num}
\bibliography{snoreFinal}

\begin{thebibliography}{10}
\expandafter\ifx\csname url\endcsname\relax
  \def\url#1{\texttt{#1}}\fi
\expandafter\ifx\csname urlprefix\endcsname\relax\def\urlprefix{URL }\fi
\expandafter\ifx\csname href\endcsname\relax
  \def\href#1#2{#2} \def\path#1{#1}\fi

\bibitem{vandewalle-2001}
N.~Vandewalle, J.~F. Lentz, S.~Dorbolo, F.~Brisbois, Avalanches of popping
  bubbles in collapsing foams, Phys. Rev. Lett. 86 (2001) 179.

\bibitem{hurst51}
H.~E. Hurst, Long-term storage capacity of reservoirs, Trans. Am. Soc. Civ.
  Eng. 116 (1951) 770--799.

\bibitem{alencar-2001}
A.~M. Alencar, S.~V. Buldyrev, A.~Majumdar, H.~E. Stanley, B.~Suki, Avalanche
  dynamics of crackle sound, Phys. Rev. Lett. 87 (2001) 088101.

\bibitem{alencar-2003}
A.~M. Alencar, S.~V. Buldyrev, A.~Majumdar, H.~E. Stanley, B.~Suki, Relation
  between crackle sound and the perimeter growth of a cayley tree: Application
  to lung inflation, Phys. Rev. E 68 (2003) 011909.

\bibitem{ishizaki-2008}
R.~Ishizaki, T.~Shinba, G.~Mugishima, H.~Haraguchi, M.~Inoue, Time-{series}
  analysis of {sleep}–-{wake} stage of rat {EEG} using time-dependent pattern
  entropy, Physica A 387 (2008) 3145.

\bibitem{acosta-2011}
E.~{N\'u\~nez-Acosta}, C.~Lerma, M.~F. M\'arquez, M.~V. Jos\'e, Mutual
  information analysis reveals bigeminy patterns in {Andersen}--{Tawil}
  syndrome and in subjects with a history of sudden cardiac death, Physica A
  387 (2008) 3145.

\bibitem{fairbanks-1994}
D.~N.~F. Fairbanks, S.~Fujita, Snoring: An overview with historical
  perspectives, in Snoring and Obstructive Sleep Apnea, Raven Press, New York,
  1994.

\bibitem{flemons-2002}
W.~W. Flemons, Clinical practice. obstructive sleep apnea, N Engl J Med 347~(7)
  (2002) 498--504.

\bibitem{drager-2011}
R.~F. Drager, V.~Y. Polotsky, G.~Lorenzi-Filho, Obstructive sleep apnea: an
  emerging risk factor for atherosclerosis, Chest 140 (2011) 534--542.

\bibitem{ruehland-2009}
W.~R. Ruehland, P.~D. Rochford, F.~J. O'Donoghue, R.~J. Pierce, P.~Singh, A.~T.
  Thornton, The new aasm criteria for scoring hypopneas: Impact on the apnea
  hypopnea index, Sleep 32~(2) (2009) 150--157.

\bibitem{chervin-2000}
R.~D. Chervin, Sleepiness, fatigue, tired, and lack of energy in obstructive
  sleep apnea, Chest 118 (2000) 372--379.

\bibitem{mainmon-2010}
N.~Maimon, P.~J. Hanly, Does snoring intensity correlate with the severity of
  obstructive sleep apnea?, Journal of Clinical Sleep Medicine 6~(5) (2010)
  475--478.

\bibitem{ghaemmaghami-2010}
H.~Ghaemmaghami, U.~R. Abeyratne, C.~Hukins, Normal probability testing of
  snore signals for diagnosis of obstructive sleep apnea, 31st Annual
  International Conference of the IEEE (2009) 5551--5554.

\bibitem{ng-2009}
A.~K. Ng, T.~S. Koh, U.~R. Abeyratne, K.~Puvanendran, Investigation of
  obstructive sleep apnea using nonlinear mode interactions in nonstationary
  snore signals, Annals of Biomedical Engineering 37~(9) (2009) 1796--1806.

\bibitem{cavusoglu-2007}
M.~Cavusoglu, M.~Kamasak, O.~Erogul, T.~Ciloglu, Y.~Serinagaoglu, T.~Akcam, An
  efficient method for snore/nonsnore classification of sleep sounds, Physiol.
  Meas. 28 (2007) 841.

\bibitem{cavusoglu-2008}
M.~Cavusoglu, T.~Ciloglu, Y.~Serinagaoglu, M.~Kamasak, O.~Erogul, T.~Akcam,
  Investigation of sequential properties of snoring episodes for obstructive
  sleep apnoea identification, Physiol. Meas. 29 (2008) 879.

\bibitem{lee-2004}
J.~Lee, D.~Kim, I.~Kim, K.~S. Park, S.~I. Kim, Nonlinear analysis of human
  sleep {EEG} using detrended fluctuation analysis, Medical Engineering \&
  Physics 26 (2004) 773.

\bibitem{rajendra-2005}
U.~{Rajendra Acharya}, O.~Faust, N.~Kannathal, T.~Chua, S.~Laxminarayan,
  Non-linear analysis of {EEG} signals at various sleep stages, Comput. Methods
  Programs Biomed. 80 (2005) 37.

\bibitem{zhang-2009}
J.~Zhang, X.~Yang, L.~Luo, J.~Shao, C.~Zhang, J.~Ma, G.~Wang, Y.~Liu, C.~Peng,
  J.~Fang, Assessing severity of obstructive sleep apnea by fractal dimension
  sequence analysis of sleep {EEG}, Physica A 388 (2009) 4407.

\bibitem{vieira10}
A.~P. Vieira, E.~P. de~Moura, L.~L. Gon\c{c}alves, Fluctuation analyses for
  pattern classification in nondestructive materials inspection, EURASIP
  Journal of Advances in Signal Processing 2010 14 (2010) 262869.

\bibitem{feder88}
J.~Feder, Fractals, Plenum Press, New York, 1988.

\bibitem{goldberger2002}
A.~L. Goldberger, L.~A.~N. Amaral, J.~M. Hausdorff, P.~C. Ivanov, C.-K. Peng,
  H.~E. Stanley, Fractal dynamics in physiology: Alterations with disease and
  aging, Proceedings of the National Academy of Sciences of the United States
  of America 99~(Suppl 1) (2002) 2466--2472.
\newblock \href {http://dx.doi.org/10.1073/pnas.012579499}
  {\path{doi:10.1073/pnas.012579499}}.

\bibitem{hausdorff2000}
J.~M. Hausdorff, A.~Lertratanakul, M.~E. Cudkowicz, A.~L. Peterson, D.~Kaliton,
  A.~L. Goldberger, Dynamic markers of altered gait rhythm in amyotrophic
  lateral sclerosis, Journal of Applied Physiology 88~(6) (2000) 2045--2053.

\bibitem{hwa2002}
R.~C. Hwa, T.~C. Ferree, Scaling properties of fluctuations in the human
  electroencephalogram, Phys. Rev. E 66 (2002) 021901.

\bibitem{peng2002}
C.~K. Peng, J.~E. Mietus, Y.~H. Liu, C.~Lee, J.~M. Hausdorff, H.~E. Stanley,
  A.~L. Goldberger, L.~A. Lipsitz, Quantifying fractal dynamics of human
  respiration: Age and gender effects, Ann. Biomed. Eng. 30~(5) (2002)
  683--692.

\bibitem{webb02}
A.~R. Webb, Statistical Pattern Recognition, 2nd ed., John Wiley \& Sons, West
  Sussex, 2002.

\bibitem{2010-pevernagie}
D.~Pevernagie, R.~M. Aarts, M.~{De Meyer}, The acoustics of snoring, Sleep
  Medicine Reviews 14 (2010) 131--144.

\end{thebibliography}
\end{document}